\documentclass[a4paper,11pt]{article}
\usepackage[german,american]{babel}
\usepackage[a4paper,top=3.5cm,bottom=3.5cm,left=3.0cm,right=3.0cm,marginparwidth=1.75cm]{geometry}
\usepackage{amsmath}
\usepackage{amsfonts}
\usepackage{amssymb}
\usepackage{graphicx}
\usepackage[font=small,labelfont=bf]{caption}
\usepackage{subcaption}
\usepackage{float}
\usepackage[OMLmathsfit]{isomath}
\usepackage{amssymb} 
\usepackage{amsxtra} 
\usepackage{bm} 
\usepackage{mathtools}
\usepackage[usenames]{color}
\usepackage{verbatim}
\usepackage[colorlinks=true, allcolors=blue]{hyperref}
\usepackage{lineno}
\usepackage{soul}

\DeclareCaptionLabelFormat{mylabel}{#1 #2\hspace{1.5ex}}
\usepackage[numbers,sort&compress]{natbib}

\definecolor{darkgreen}{rgb}{0.0,0.5,0.0}
\definecolor{BurntOrange}{rgb}{0.8,0.3,0.0}
\definecolor{mygray}{gray}{0.5}

\include{new_commands}

\providecommand{\keywords}[1]
{
  {
  \small	
  \textbf{\textit{Keywords---}} #1
}}

\title{Accelerating phase-field-based simulation via machine learning}
\author{Iman Peivaste$^{1}$, 
Nima H. Siboni$^{2}$,
Ghasem Alahyarizadeh$^{1,*}$, 
Reza Ghaderi$^{3}$, \\ 
Bob Svendsen$^{2,4}$, 
Dierk Raabe$^{2}$, 
Jaber Rezaei Mianroodi$^{2,*}$\\
        \footnotesize $^1$Faculty of Engineering, Shahid Beheshti University, Tehran, Iran\\ 
       \footnotesize $^2$Microstructure Physics and Alloy Design, \\ 
        \footnotesize Max-Planck-Institut f\"ur Eisenforschung, D\"usseldorf, Germany \\
        \footnotesize $^3$Department of Electrical Engineering, Shahid Beheshti University, Tehran, Iran \\
        \footnotesize $^4$Material Mechanics, RWTH Aachen University, 52062 Aachen, Germany \\
        \footnotesize $^*$Corresponding Authors: j.mianroodi@mpie.de, g\_alahyarizadeh@sbu.ac.ir
}

\begin{document}

\maketitle
\begin{abstract} 
\noindent 
Phase-field-based models have become common in material science, mechanics, physics, biology, chemistry, and engineering for the simulation of microstructure evolution. Yet, they suffer from the drawback of being computationally very costly when applied to large, complex systems. To reduce such computational costs, a Unet-based artificial neural network is developed as a surrogate model in the current work. Training input for this network is obtained from the results of the numerical solution of initial-boundary-value problems (IBVPs) based on the Fan-Chen model for grain microstructure evolution. In particular, about 250 different simulations with varying initial order parameters are carried out and 200 frames of the time evolution of the phase fields are stored for each simulation. The network is trained with 90\% of this data, taking the $i$-th frame of a simulation, i.e. order parameter field, as input, and producing the $(i+1)$-th frame as the output. Evaluation of the network is carried out with a test dataset consisting of 2200 microstructures based on different configurations than originally used for training. The trained network is applied recursively on initial order parameters to calculate the time evolution of the phase fields. The results are compared to the ones obtained from the conventional numerical solution in terms of the errors in order parameters and the system's free energy. The resulting order parameter error averaged over all points and all simulation cases is 0.005 and the relative error in the total free energy in all simulation boxes does not exceed 1\%.
\end{abstract}

\section{Introduction}

Over the past decades, phase-field (PF) approaches have become one of the 
most common methods for computational modeling and simulation of microstructure 
evolution 
in materials science and engineering as well as several other 
disciplines \cite{Mannor2005,Pf_evol_book,
biner2017overview,boettinger2002phase, steinbach2009phase, porter2009phase, chen2002phase}. 
In the case of metallic alloy systems, PF-based models have 
been developed for a number of phenomena including solidification, 
structural phase transformation, chemical species transport, 
precipitation, defects, and plasticity, or recrystallization, mostly aiming at the prediction of microstructure evolution at the mesoscale
\cite{li2017review, millett2011application,
tonks2018apply,ambati2015review, cervera2021comparative, steinbach2013solidification,
mamivand2013review,bui2021review, tonks2019phase, chen2002phase, singer2008phase}.

In PF-based models for metallic alloys, the phases, their associated microstructures, and chemical composition are described by PFs that vary smoothly 
across phase interfaces. The energy landscape of the material, 
from which local driving forces are derived, is parameterized by the 
PFs, or the order parameters in the case of non-conservative phase fields.
In the case of metallic alloys, PF-based models are 
either of Cahn-Hilliard (CH) type \cite{cahn1958free, cahn1961spinodal} 
for conservative phase fields (e.g., chemical concentration) 
or Allen-Cahn (AC) \cite{allen1972ground, allen1973correction} 
type for non-conservative phase fields (e.g., structural order parameters). 

Although PF-based models can describe microstructure evolution 
in a broad range of physical, biological, and chemical systems, 
the corresponding computational efforts required to solve them are often prohibitively expensive. 
The reasons for this include (i) the fine spatial and temporal 
discretization required for a converged (i.e., physical) solution 
of the corresponding initial-boundary-value problems (IBVPs) 
and (ii) iteration-based and often staggered numerical schemes to solve the underlying partial differential equations. Particularly 
in the case of coupled multi-physics models (where for instance structural, mechanical, and chemical fields must be considered), these are principle 
obstacles to the simulation of microstructure evolution, especially 
in systems much larger than the currently tackled microstructural lengthscales. 
In the multiphysics case, multiple material lengthscales 
need to be resolved in the same spatial domain during 
simulation based on the numerical solution of the corresponding coupled 
IBVPs. Such multiphysics cases in materials science and engineering include 
Li-ion batteries, hydrogen embrittlement, stress corrosion cracking, 
iron-ore reduction, or solidification, which involve a non-linear interplay 
of microstructure, chemistry, and mechanics. 
Finding ways to reduce the computational effort required for the PF-based 
computational modeling and simulation of such cases is essential for the extension of the PF approach to even larger and more complex systems. 
The core idea behind this is not simply to be able to deal with larger system sizes but to properly reflect the different coherence lengths involved in the respective fields (such as interface density, dislocation density, grain size, diffusion length,  etc.). Also, the behavior of certain complex systems is often hidden behind the interplay of structure, mechanics, and chemistry. This means that the coupling that enters through multi-physics simulations and the system size becomes integral tasks of the proper formulation of the actual scientific problem addressed. 

A number of steps have been taken to tackle this challenge; for example, 
developing more efficient numerical techniques \cite{Cheng2015FastAS, Tang2019ONEN}, 
parallel computing, or employing graphics processing units (GPU). 
Despite the success of these, the ever-increasing computational complexity, 
required system size, and number of variables, require even further 
increase in computational speed 
\cite{alizadeh2020managing, mianroodi2021teaching, Li2021FourierNO}. 
As a result, alternative approaches such as machine learning (ML) have received a lot of attention in recent years.

ML offers a range of powerful techniques, some of which have 
been applied in
computational materials science 
\cite{huang2021artificial, ward2018strategies, schmidt2019recent}. 
Currently, many studies are focusing on applying ML in different areas  
of materials science \cite{schmidt2019recent, e40208aa0bc14c2a9d37d4c4180e16fc}. 
One recent application of ML in this regard is the development of so-called 
surrogate computational models based on trained artificial neural networks 
(tANNs). 
For instance, Mianroodi et al.~\cite{mianroodi2021teaching} 
developed an ANN for surrogate computational modeling of the stress field 
in a grain microstructure consisting of elastoplastic single crystals. 
They demonstrated that the calculation of the stress field using the 
tANN-based surrogate model is up to 8000 times faster than the 
spectral-based numerical solution of the corresponding periodic IBVP, even for grain microstructures not included in the training data set.
In a different context, Nyshadham et al. \cite{nyshadham2019machine} 
developed a fast surrogate model to estimate the properties of materials 
such as elastic constants, enthalpy, and band-gap, with the accuracy 
of \textit{ab initio} methods. In addition, surrogate models have 
also been employed in up-scaling of multi-scale systems 
(see \cite{peng2021multiscale, wang2018multiscale, mianroodi2021lossless}).

In the case of PF-based modeling, several attempts have been made 
to develop surrogate models for the evolution of 
chemical composition and structural order parameter fields  
based on spatio-temporal pattern correlation.
Recently, Montes de Oca Zapiain et al.~\cite{de2021accelerating} 
presented a new data-driven surrogate model capable of predicting 
microstructure evolution in less than a second. They used a long-short-term memory network 
trained with results from the numerical solution of a physical IBVP.
Comparison of results from their surrogate model for spinodal 
decomposition with those from the solution of the IBVP show 95\% agreement. 
Zhang et al. \cite{zhang2020machine} developed a surrogate 
model based on deep neural networks (DNNs) and convolutional 
neural networks (CNNs) trained using free energy fluctuations and used 
this to calculate the stress in their microstructure. 
They showed that both CNN-based and DNN-based neural network 
architectures are performing equally in capturing 
microstructure features and predicting free energy evolution. 
Teichert and Garikipati \cite{teichert2019machine} carried out 
similar work in which DNNs were 
trained with free energy results from the numerical solution of a 
PF-model-based IBVP 
and employed in the surrogate modeling of precipitate morphology. 
In another study, Hermen et al.~\cite{herman2020data} developed a 
trained ANN based on results for chemical vapor deposition and applied 
this to the surrogate modeling of a corresponding microstructure. 
Similar surrogate models have been developed in 
\cite{jiang2019fast, latypov2019bisque,yabansu2019application}. 

Evaluation of surrogate models is often based on three criteria 
\cite{alizadeh2020managing}:~
(i) size of the training dataset, 
(ii) time needed to train a neural network and obtain results, 
and (iii) model accuracy. 
In particular, (i) and (ii) are related to the efficiency 
of the training process. As usual, there is a trade-off between 
efficiency and accuracy, and so a need for further optimization, i.e., to reduce 
(i) and (ii) while maintaining (iii). 
In other words, a need exists to develop fast and accurate 
surrogate models based on training datasets of minimal size 
and training effort. 

In this study, we develop a surrogate model trained with results from the numerical solution of IBVPs based on the Fan-Chen model for grain growth \cite{fan1997computer} and semi-implicit spectral methods. The resulting model is capable of calculating the change in order-parameter fields that represent the microstructure evolution at many timesteps forward. Compared to other approaches, the method suggested here is a point-wise neural network that requires a considerably smaller dataset (a few hundred microstructure evolution paths) compared to the other neural network architectures. A U-net-based ANN architecture is used in this work. Since the free energy is a functional of the order parameters, it is used as an accuracy measure. 
Following the introduction, the work begins in Section \ref{model} with a 
summary of the Fan-Chen model and its numerical implementation. 
Based on this, Section \ref{sec_data_prep} discusses the training dataset. In 
Section \ref{sec_network}, the ANN architecture, input and output, and training 
are discussed. Evaluation of the resulting tANN-based surrogate model 
is carried out in Section \ref{result} with the help of a number of comparison 
measures (e.g., the free energy) and benchmark cases for microstructure 
evolution. Besides on accuracy, emphasis is placed here on the 
increase in computational speed achieved. 
Finally, in Section \ref{conc}, we draw conclusions and discuss future 
directions for the current work. 

\section{Fan-Chen model and numerical implementation}
\label{model}
\subsection{Model}

Grain growth is an important and often property-critical phenomenon in metallurgy \cite{Flint2019PhaseFieldSO}. 
In conventional grain growth, grain boundaries move due to local capillary forces acting on them, minimizing the system's total free energy, stemming in this case from the interfacial energy. Grain growth is of paramount importance owing to its impact on material properties \cite{Flint2019PREDICTIONOG, ZOLLNER2016}. Consequently, it has been studied substantially from atomistic to continuum scales \cite{MIODOWNIK2002125, najafkhani2021recent, ZOLLNER2016, louat1974theory}. At the mesoscale, several models have been proposed to simulate this phenomenon by using the PF method. One of the well-known and commonly used models is the Fan-Chen model \cite{fan1997computer}. In this model, each connected domain of the same crystallographic orientation, referred to as a grain, is represented by one specific non-conserved structural order parameter (or field variable) $\eta\textsubscript{i}$ which is 1 inside the $i\textsubscript{th}$ grain and 0 in all other grains. For example, in grain number 2,  $\eta\textsubscript{2}$ assumes the value  1, and in all other grains it is 0,  changing its value smoothly across grain boundaries. Interested readers are referred to \cite{fan1997computer} for more details about the model. The evolution of order parameters ($\eta\textsubscript{i}$) takes place by the non-conservative Allen–Cahn equation (or, respectively, the time-dependent Ginzburg-Landau equation)
\begin{equation}
    \frac{\partial \eta\textsubscript{i}(r,t)}{\partial t} 
    = -L_i \frac{\delta F}{\delta\eta_i(r,t)}
    \,,\quad i = 1,2,...,p\,,
    \label{eq_AC}
\end{equation}
where $p$ is the number of grains, $L_i$ is the 
mobility, $t$ is time, and 
\begin{equation}
    F = \int \Biggl[\!f(\eta_1, \eta_2, ..., \eta_p)+ \sum\limits_{i}^{p} \frac{\kappa_i}{2}\,| \nabla \eta_i^2| \Biggr]  \,dv
    \label{eq_freeEn}
\end{equation}
is the free energy functional. In \eqref{eq_freeEn}, $f$ is the local free energy, and $\kappa_i$ are the gradient energy coefficients, which are taken as constant here.
Taking the functional derivative of \eqref{eq_freeEn} and combining it with \eqref{eq_AC}, we get
\begin{equation}
    \frac{\partial \eta_i}{\partial t} 
    = -L \frac{\partial f}{\partial \eta_i} + L \kappa \nabla^2 \eta_i.
    \label{eq_AC2}
\end{equation}
The local free energy is written as
\begin{equation}
   f(\eta_1, \eta_2, ..., \eta_p)
   = \sum\limits_{i}^{p} \biggl( -\frac{\alpha}{2} \eta_i^2+\frac{\beta}{4} \eta_i^4 \biggr)+\gamma \sum\limits_{i}^{p}\sum\limits_{i\neq j}^{p} \eta_i^2 \eta_j^2,
   \label{eq_free_en}
\end{equation}
where $\alpha$, $\beta$ and $\gamma$ are positive constants.
 
\subsection{Numerical implementation}
\label{numerical implementation}

In this work, the spectral method, in particular, the Fast Fourier Transform (FFT), is chosen as the numerical tool to solve the PDE in \eqref{eq_AC} and generate high-fidelity reference data for training and testing the artificial neural network. In this method, quantities are transformed from the real space to the reciprocal space as
\begin{equation}
    \hat{\eta} (k, t) 
    = \int \limits_{- \infty}^{\infty} \eta (r, t) \exp (ik,r) \,dx \, ,
\end{equation} 
where $k$ is the wave vector. The spatial derivatives in Fourier space are 
given by
\begin{equation}
    \frac{\partial^n \hat{\eta}(k,t)}{\partial x^n} 
    = (\imath k)^n\,\hat{\eta} (k, t)
    \,,\quad\imath=\sqrt{-1}
\,.
\end{equation} 
Taking the Fourier transform of the two sides of \eqref{eq_AC2}, the reciprocally-discrete equation is
\begin{equation}
    \frac{\partial \hat{\eta}_i(k,t) }{\partial t} 
    = -L \widehat{\biggl[ \frac{\partial f}{\partial \eta_i (r, t)} \biggr]_k}-k^2 L \kappa \hat{\eta} (k, t)
    \,.
    \label{eq_ACFFT}
\end{equation}
Note that the quantity inside the bracket $[]_k$ above must be transformed into Fourier space. $k$ consists of $x$ and $y$ components with a magnitude of $\sqrt {k_x^2+k_y^2}$. The semi-implicit form of
\eqref{eq_ACFFT} is
\begin{equation}
    \eta_i^{n+1} (k, t) = \dfrac{\eta_i^{n} (k, t) 
    - \Delta t\,L 
    \biggl[\widehat{\dfrac{\partial f}{\partial \eta_i (r, t)}}\biggr]_k^n }
    {1+\Delta t\,L \kappa k^2}\,,
    \label{eq_SemiImp}
\end{equation}
where 
$\Delta t$ 
is the timestep size. 
It is worthwhile to mention that for the forward and the backward transform 
into and from Fourier space, respectively, fft and ifft functions from 
the Numpy \cite{harris2020array} package in python were used.

\section{Data preparation} 
\label{sec_data_prep}

All parameters are in non-dimensional form.
The 2D simulation domain is a square grid with 128 points in $x$ and $y$ 
directions ($N_x$ = 128 and $N_y$ = 128). 
A grid point spacing of $\Delta x=\Delta y=1.0$, 
and a timestep size of $\Delta t = 8 \times 10^{-3}$, are employed for 
the numerical solution of Equation \eqref{eq_SemiImp}. In addition, values of 
20 and 0.3 are adopted for the relaxation coefficient $L_i$ and the 
gradient energy coefficient $\kappa$, respectively. Likewise, 
$\alpha$, $\beta$ and $\gamma$ in Equation \eqref{eq_free_en} 
are all assumed equal to 1.0.  
The choice of these parameters ensures a numerically well-resolved interface as well as reaching equilibrium before the end of the simulations.

In this study, for simplicity, we only consider binary grain structures. 
This means that we use only two types of crystals (grains) with 
homogeneous internal structure, i.e. they have a constant value 
of the structure variable in their interior, so that a grain is constituted by a
homogeneous domain that assumes a value of 1 for one of the two order parameters. 
Therefore, there are only two order parameters used to describe the microstructure 
in this domain ($\eta_1$ and $\eta_2$). $\eta_1$ takes a value of 1 in the first 
grain and 0 in the second one. Likewise, $\eta_2$ takes the value of 0 in the first
grain and 1 in the second grain. The regions where these order parameters assume a
value between 0 and 1 are the interface regions.
Interface regions contribute an extra energy term to the system, resulting in a driving force for the evolution of the grain boundaries.

\begin{figure}[H]
    \centering
    \includegraphics[width=0.6\textwidth]{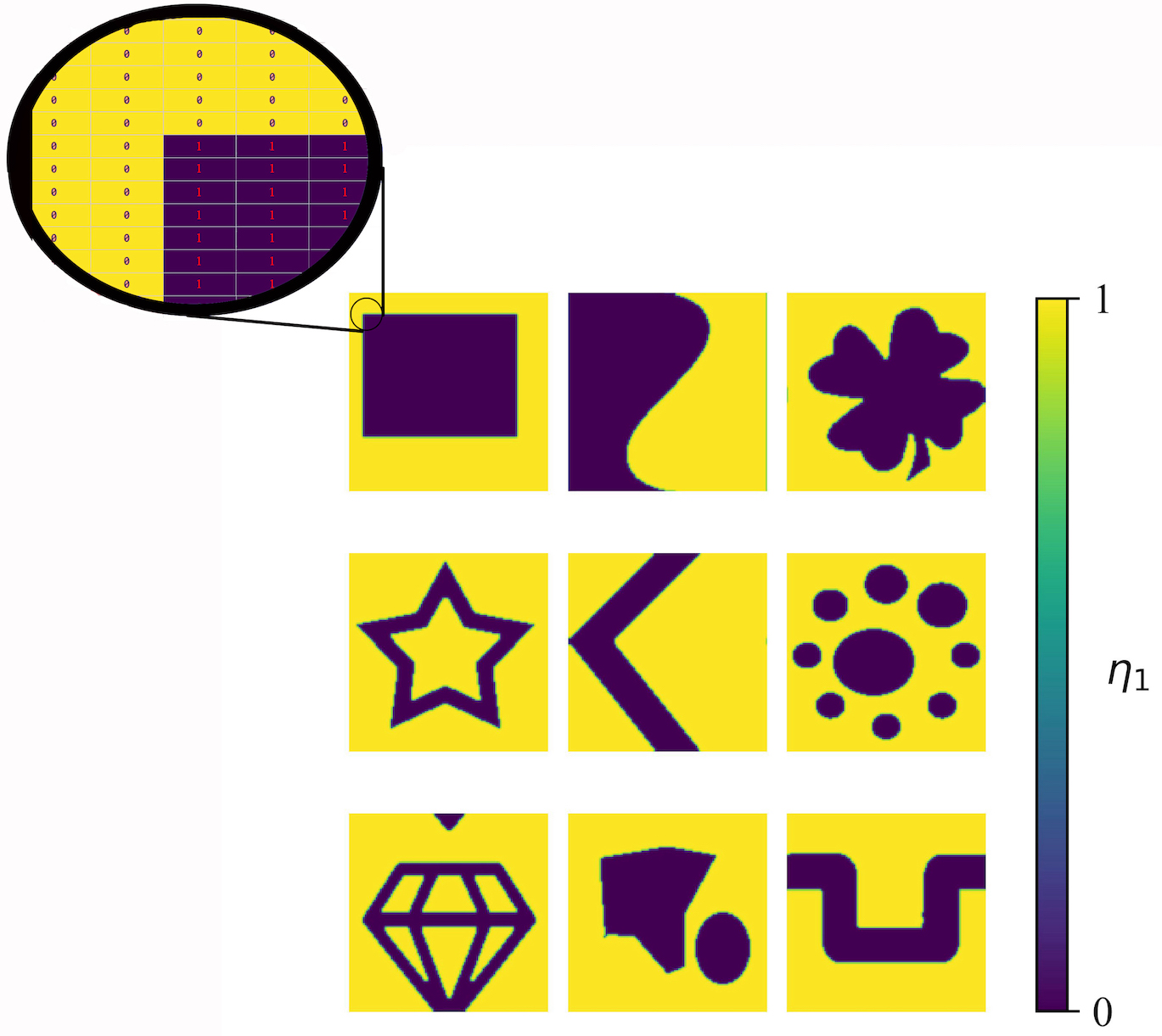}
    \caption{\textbf{Examples of the initial phase-field configurations}. $\eta_1$ is set to 0 and 1 inside and outside the grain, respectively.}
    \captionsetup{justification=centering, margin=2cm}
    \label{fig:inittial}
\end{figure}

\begin{figure}[H]
    \centering
    \includegraphics[width=0.7\textwidth]{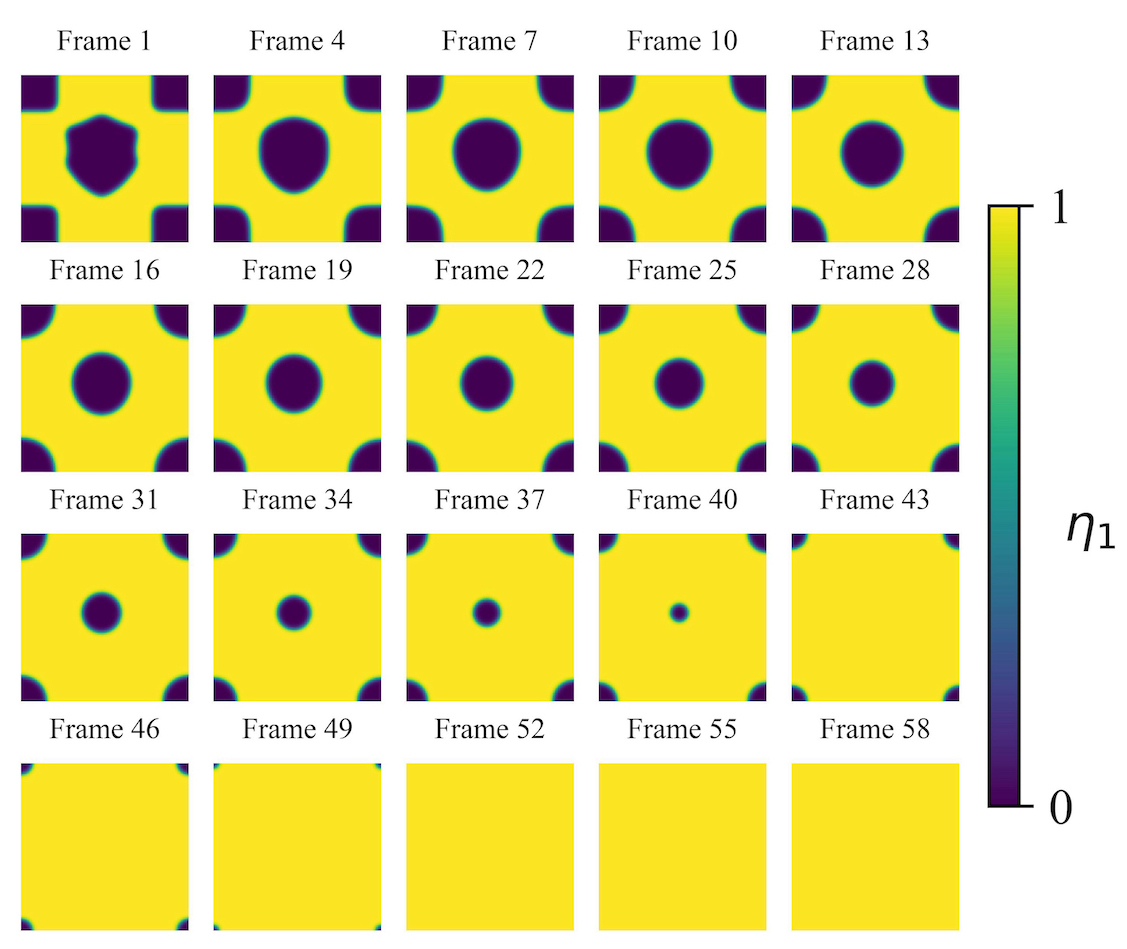}
    \caption{\textbf{Evolution of an exemplary microstructure} shown in terms of the spatial distribution of the values of the two order parameters $\eta_1$. Frame 1 is the first solution stored after the artificially sharp initial interfaces of the start configuration have been relaxed. After the $50^{th}$ frame, the outer grain eliminates the inner grain, and the microstructure reaches a stable form. Note that each frame here represents the progress after a sequence of 50 individual timesteps.}
    \label{fig:20_frame}
\end{figure}

Initial order parameter field values in the simulation cell 
determining initial grain shapes are chosen randomly from a set of images. 
Fig. \ref{fig:inittial} depicts 9 different examples of initial
configurations of $\eta_1$. 
Since the system consists of only two grains, the second order parameter would be $\eta_2 = 1- \eta_1$.
The evolution of each configuration is simulated for 10,000 timesteps, and the
solution of \eqref{eq_SemiImp} 
for \(i=1\) (\(\eta_{2}=1-\eta_{1}\)) 
is stored every 50 timesteps. As a result, the temporal evolution of each initial configuration is captured in the form of 200 subsequent frames. In 
other words, at the end of each simulation, 200 consecutive solution snapshots are saved, each
containing the spatial distribution of the two order parameters in the 2D domain. 
We eliminate the first snapshot, as it has sharp interfaces and it is far from equilibrium. An example of the full evolution of one microstructure 
is shown in Fig. \ref{fig:20_frame}. 
In the end, the 250 initial configurations generate 500,000 frames, each containing the evolving grain growth patterns produced by two order parameters, $\eta_1$, and $\eta_2$. 
As shown for example in Fig. \ref{fig:20_frame}, one grain grows 
(yellow in the figure) at the expense of the other during microstructure 
evolution, resulting in a homogeneous single crystal. In particular, due 
to concavity, the capillary driving force drives shrinkage of the unstable 
grain which completely disappears after the $50^{th}$ frame (or 2500 timesteps).
In such cases, only the first two homogeneous solutions have been stored. 
Our goal is to train the network to
reproduce the time-dependent behavior of the microstructure.

\section{Neural network architecture and training}
\label{sec_network}
Most physical phenomena in materials science can be modeled through 
solving PDE-based mathematical formulations, hence, any method 
accelerating their solution is important in this field.
In this context, trained ANNs as surrogate models to reproduce the numerical solution of PDEs are very promising. 

In the current work, results from the solution of 
the IBVP discussed in the previous section are used to train an ANN directly, 
resulting in a tANN-based surrogate model. 
Our goal is to develop a tANN to reproduce 
the time evolution of the phase-field order parameter 
based on the numerical solution of a periodic IBVP based on 
the Fan-Chen model for grain growth as discussed above. 
To this end, a "U-net" architecture \cite{ronneberger2015u} 
is adopted and modified for this purpose. U-net has been 
mainly used for classification and segmentation in computer 
vision. 
Recently, however, 
Mianroodi et al. \cite{mianroodi2021lossless} trained a 
U-net-based ANN for surrogate modeling of the stress field 
in polycrystalline microstructures. 
In this work, we show that a similar U-net can effectively reproduce 
results from the spatio-temporal numerical solution of an IBVP 
based on the Fan-Chen model. 

\subsection{Neural network architecture}

\begin{figure}[H]
    \centering
    \includegraphics[width=0.9\textwidth]{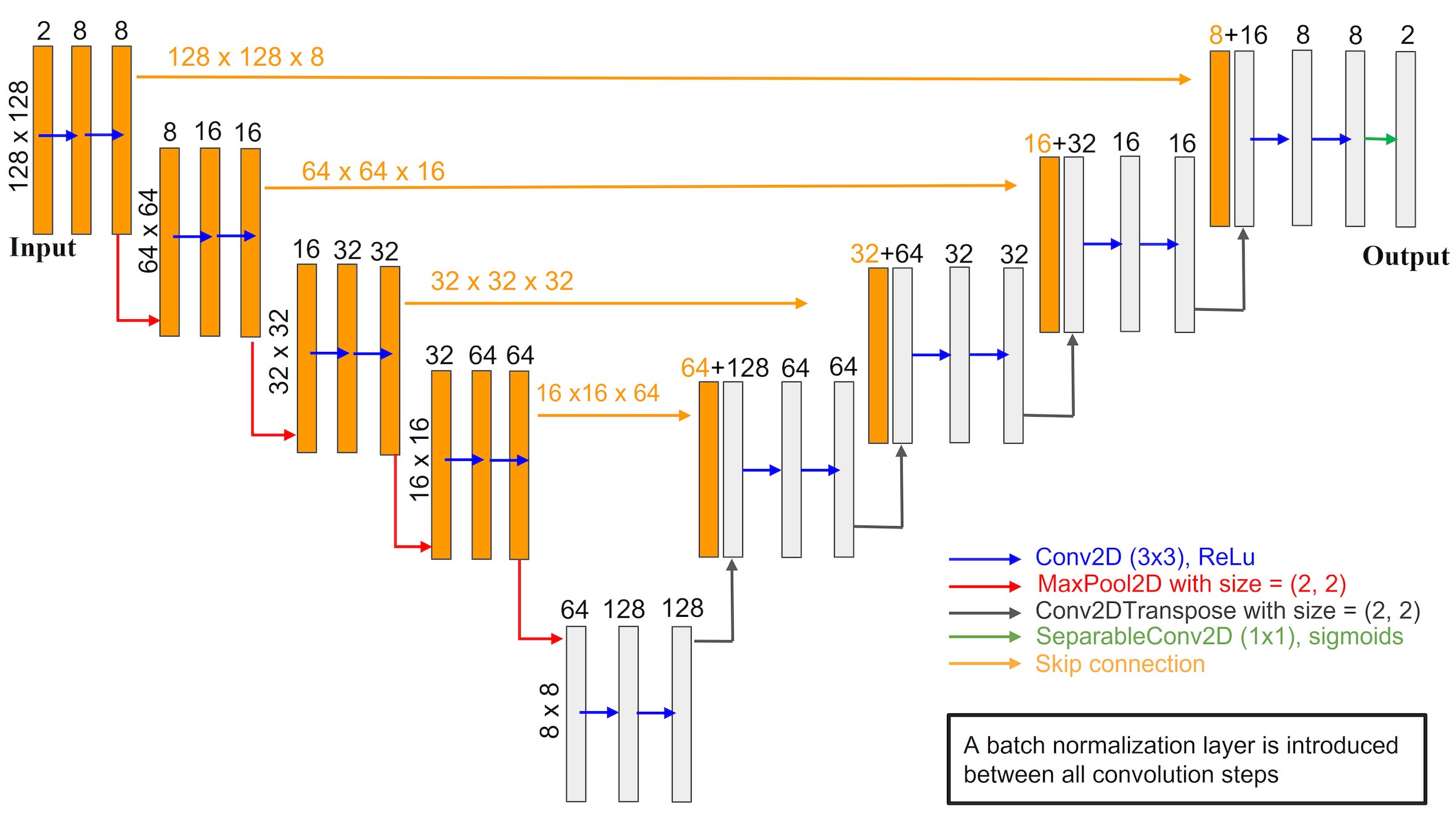}
    \caption{\textbf{U-net architecture employed in this study.} 
    The network has been constructed using TensorFlow 
\cite{tensorflow2015-whitepaper} in conjunction with the 
Keras \cite{chollet2015keras} library. 
The network's encoder and decoder parts consist of four stages. 
Each encoder and decoder stage contains two Conv2D layers (blue arrows). The encoder part consists of downsampling (by a max-pool operation, red arrow) while the decoder part conducts the upsampling (by Conv2DTranspose operation, grey arrows).
Each part has the same number of filters, and a Rectified Linear Unit (ReLU) 
for activation, except for the last one, which uses a sigmoid activation function (green arrow). 
A 2D max-pooling operation compresses the data between each stage in 
the encoding part (red arrows), and expands it between each stage in 
the decoding part (grey arrows). Skip connections (yellow arrows) feed 
the output on one stage as the input to the corresponding stage 
in the decoding part. See text for further discussion.\newline
}
\label{fig:U_net}
\end{figure}

The input to the network in the current context consists of values of 
order parameter fields for each grain in the microstructure. Since these 
are naturally dimensionless and take values between 0 and 1, they are 
ideally suited as training information. For simplicity, attention is 
limited here to 2 dimensions and a 2-grain or 2-phase microstructure. 
As such, the input consists of values for two order parameters 
$\eta_1$ and $\eta_2$ at each node/pixel in \(N_{x}\times N_{y}\) 
phase-field simulation domain (with \(N_{x,y}=128\)) discussed above, 
resulting in the \(\lbrack 128,128,2\rbrack\) input format shown on the 
left in Fig.~\ref{fig:U_net}. 
In each of the 4 encoder stages (Fig. \ref{fig:U_net}, left side, orange), 
the data are processed by two convolution layers. 
The padding for each convolution layer is considered the "same", 
so the dimension of any output data is the same as the input data. 
Convolution layers are based on kernels that are essentially simple filters 
that identify local pattern features in the training data. 
They are a core tool in such network architectures for pattern correlation 
via sequentially applied algebraic matrix operations. Stacking convolution 
layers in each stage of the encoder facilitates the correlation of important 
features in the input data. 
Note that the kernel of the convolution layers has a fixed size (in our case the kernel is a $3\times3$ matrix). The components of the kernels are optimized during the training procedure.
The activation function used in our network is a Rectified Linear Unit (ReLU).
Batch normalization is employed between all convolution steps to compress the data. 
Downsampling is based on a Max Pooling operation with a stride size of two 
and a pool size of $2 \times 2$. 
As shown in Fig.~\ref{fig:U_net}, the input \(\lbrack 128,128,2\rbrack\) 
is compressed in the encoder to \(\lbrack 8,8,64\rbrack\). 
This is then input into the decoder and expanded through 4 decoder 
stages. The skip connections concatenate the output of the corresponding 
encoder and decoder stages (orange arrows). In the final stage on 
the right, the resulting data is transformed into the format 
\(\lbrack 128,128,2\rbrack\) of the input data.
Note that we use the terminology adopted by TensorFlow \cite{tensorflow2015-whitepaper} for the different operations in the network, such as Conv2D (generate convolution kernel), MaxPool2D (downsamples data), and Conv2DTranspose (deconvolution that upsamples data).

\subsection{Network input and output}

In Section \ref{sec_data_prep}, the generation of the dataset has been explained. 
250 initial configurations of different spatial patterns (grain microstructures) 
determined by the order parameter fields in a square domain discretized 
by 128\(\times\)128 pixels are employed as initial conditions to solve the 
initial-boundary-value problem (IBVP) based on the Fan-Chen model and 
Eq.~\eqref{eq_SemiImp} for 10,000 timesteps. The corresponding 
solutions are saved every 50 timesteps as arrays. 
Hence, the size and shape of the datasets for two order parameters 
is $\lbrack 250, 200, 128, 128, 2\rbrack$. 
Note that the initial order parameter patterns contain in part random shapes with sharp interfaces, i.e. the interface initially has zero thickness. Therefore, the time evolution of the order parameters in the few first steps of the solution is much faster, due to the large driving force from the gradient term, than during the rest of the simulations. Since we are not interested in sharp interface scenarios in this work, we removed the first frames from all simulations
(corresponding to timestep 50) from the training data set.
Therefore, the dimension of the dataset becomes 
$\left[250, 199, 128, 128, 2 \right]$. In some simulation 
cases, one of the grains will shrink and disappear depending on the 
initial conditions. Once a grain has been annihilated (overgrown), the solution will not evolve anymore as we do not include 
any nucleation process in this work. Thus, simulation 
frames with homogeneous order parameter distributions, representing a perfect single crystal, are removed from the database. After filtering such cases, around 22000 
microstructures remain in the dataset. 

\begin{figure}[H]
    \centering
    \includegraphics[width=0.6\textwidth]{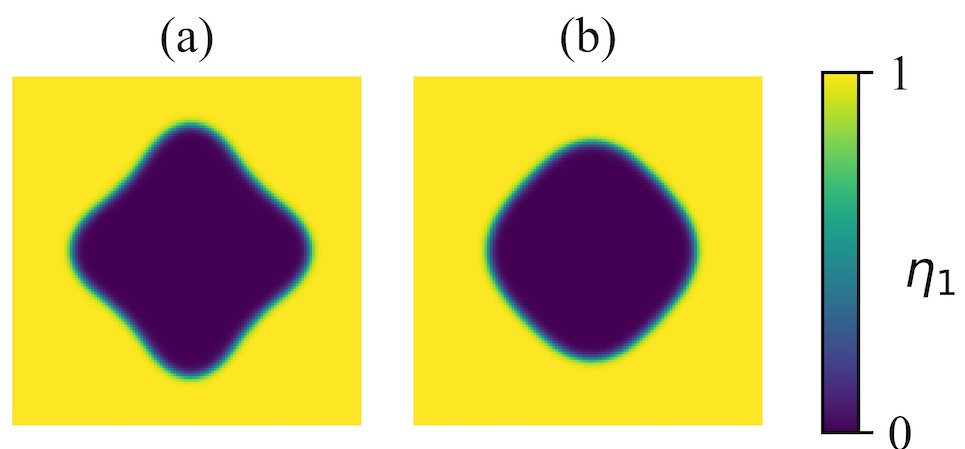}
    \caption{\textbf{An example of ANN input and output.} 
    (a) $\eta_{1}^{n}$ as input at timestep \(n\) and (b) 
    $\eta_{1}^{n+500}$ as output 500 timesteps later.}
    \label{fig:feature and label}
\end{figure}

The goal is to train the ANN to output results for the phase fields and microstructure for a given input that are close to those that would be obtained from a numerical solution of the corresponding IBVP based on the Fan-Chen model. 
An example of such a typical input
and output scenario is shown in Fig. \ref{fig:feature and label}.
Here, order parameter values for timestep \(n\) are input to the ANN, and order parameter results for timestep \(n+500\) are 
output. 

\subsection{Network training}

\begin{figure}[H]
    \centering
    \includegraphics[width=0.85\textwidth]{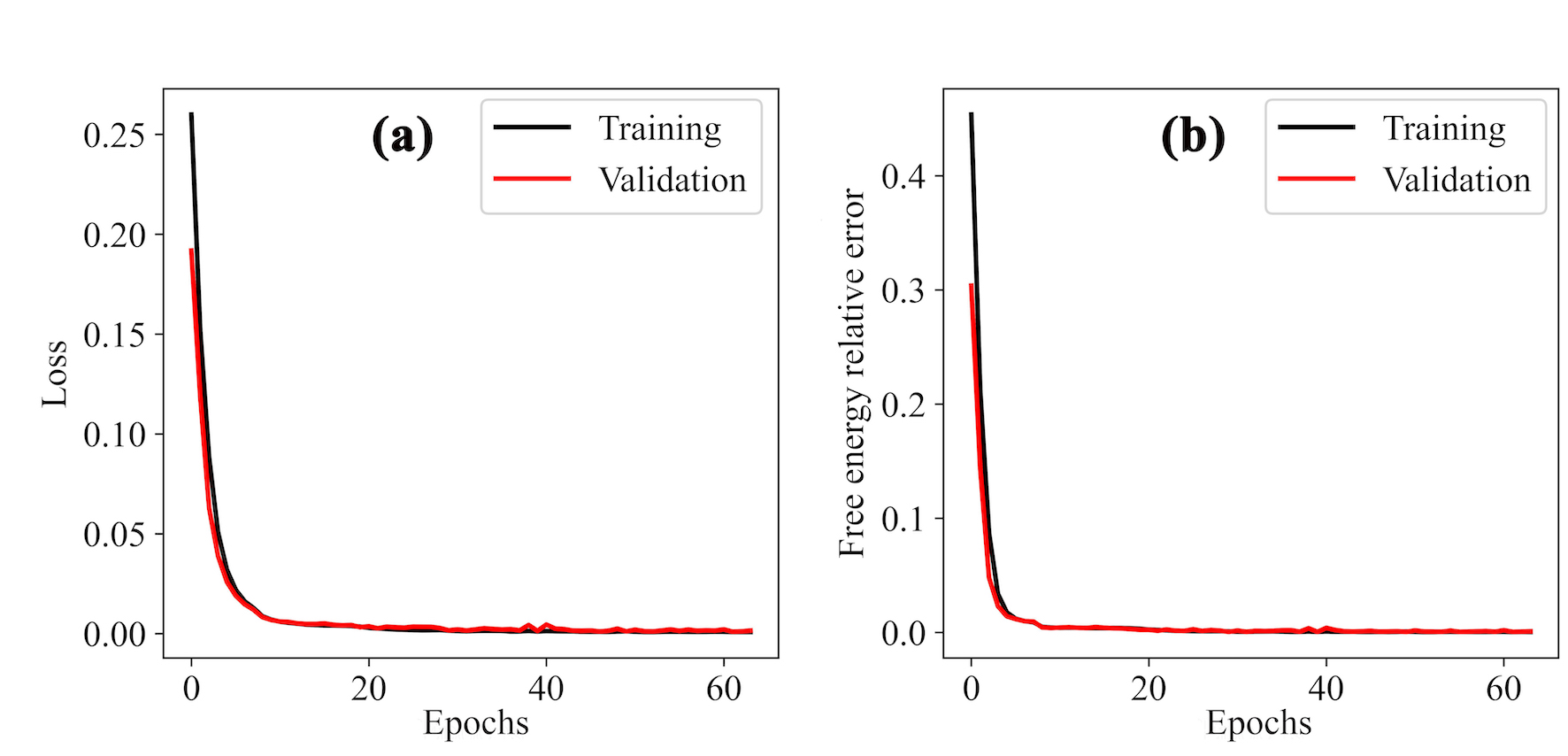}
    \caption{Value of the loss function (a) and relative error in free energy (b) 
    as a function of epoch during the training of the network.}
    \label{fig:Loss}
\end{figure}

In training and testing ANN, a dataset is typically divided into three parts of (i) training, (ii) validation, and (iii) test datasets. The training dataset is used in optimizing the trainable parameters of the network, while the performance of the network on the validation dataset is monitored during the training to detect over / underfitting. The test dataset is kept separate and only used to evaluate the accuracy of the network after training. In this work, 90\% of the entire dataset (randomly selected) has been assigned to training and the remaining 10\% to testing. 10\% of the training dataset was also considered for validating the neural network.
Note that there is a terminology discrepancy between the field of constrained optimization (CO) and machine learning which is explained in the discussions. Here we adopt the common terminology in the field of machine learning.
The total number of trainable parameters in the neural network is 487,226.
As usual, training is based on minimization of the corresponding 
objective or "loss" function, which in this work is 
the mean absolute 
error (MAE) between the results from the numerical solution ($\eta_i^{NS}$) of the IBVP 
based on the Fan-Chen model and the corresponding output of the neural network ($\eta_i^{NS}$)
\begin{equation}
    MAE = \frac{1}{m}\sum_{j=1}^{m}\sum_{i=1}^{2}\left | \left ( \eta_i^{NS} -\eta_i^{NN} \right ) \right |,
\end{equation}
where $m$ is the number of all microstructures in the dataset.
Monitoring the value of the loss function during network training is a common way to evaluate training. 
In Fig. \ref{fig:Loss} (a), the training and validation loss (based on 
the order parameters) as a function of epochs are reported. 

Based on the results in this figure, no sign of over-fitting or under-fitting is observed. 
Besides the monitored evaluation of the loss function, the quality of training is 
also examined with respect to the relative error in the free energy 
displayed in Fig.~\ref{fig:Loss} (b). 
This error is between 0.001 and 0.002 for both, training and validation. 
This means that the neural network produces a total free energy value for the respective microstructure at the next desired timestep with high precision. 

The ADAM optimizer \cite{Kingma2015AdamAM} is 
employed as the stochastic gradient descent algorithm with the 
learning rate of $5.0\times 10^{-4}$. The other parameters of the 
ADAM optimizer are as follows: $\beta_1= 0.9$, $\beta_2= 0.999$ 
and $\epsilon = 10^{-7}$. 
The neural network is trained for 64 
epochs and a batch size of 32. These are found to be the optimal training parameters (hyperparameters) in our case, resulting in the smallest values of the loss function after training. 
The free energy (which quantifies all interface energy contributions in the Fan-Chen model, see \eqref{eq_free_en}) 
is defined as an important metric for evaluating the neural network. 
The derivatives of the 
free energy with respect to the order parameters are the driving force for the evolution of the phase fields. 
After each epoch, the network compares the mean 
relative error between the results from the numerical solution of the Fan-Chen model 
and the one calculated by 
the neural network. The training takes about 30 minutes with a GeForce 
GTX 1060 GPU (compute capabilities  = 6.1).

\section{Results and discussion}
\label{result}

In the first evaluation, we investigate the effect of the timesteps 
on the neural network's performance. For this purpose, we train the 
same network three times to reproduce the evolving patterns of the 
order parameters after 100, 500, and 1000 timesteps. Next, we create 
new microstructures different from those used for training to 
investigate the network's performance on unseen microstructures. 
Finally, we evaluate the neural network's capability in determining 
the entire time series. In this case, the network's output is used 
again as the input for calculating results for several timesteps. 

\subsection{Neural network output for one frame forward}

\begin{figure}[H]
    \centering
    \includegraphics[width=0.7\textwidth]{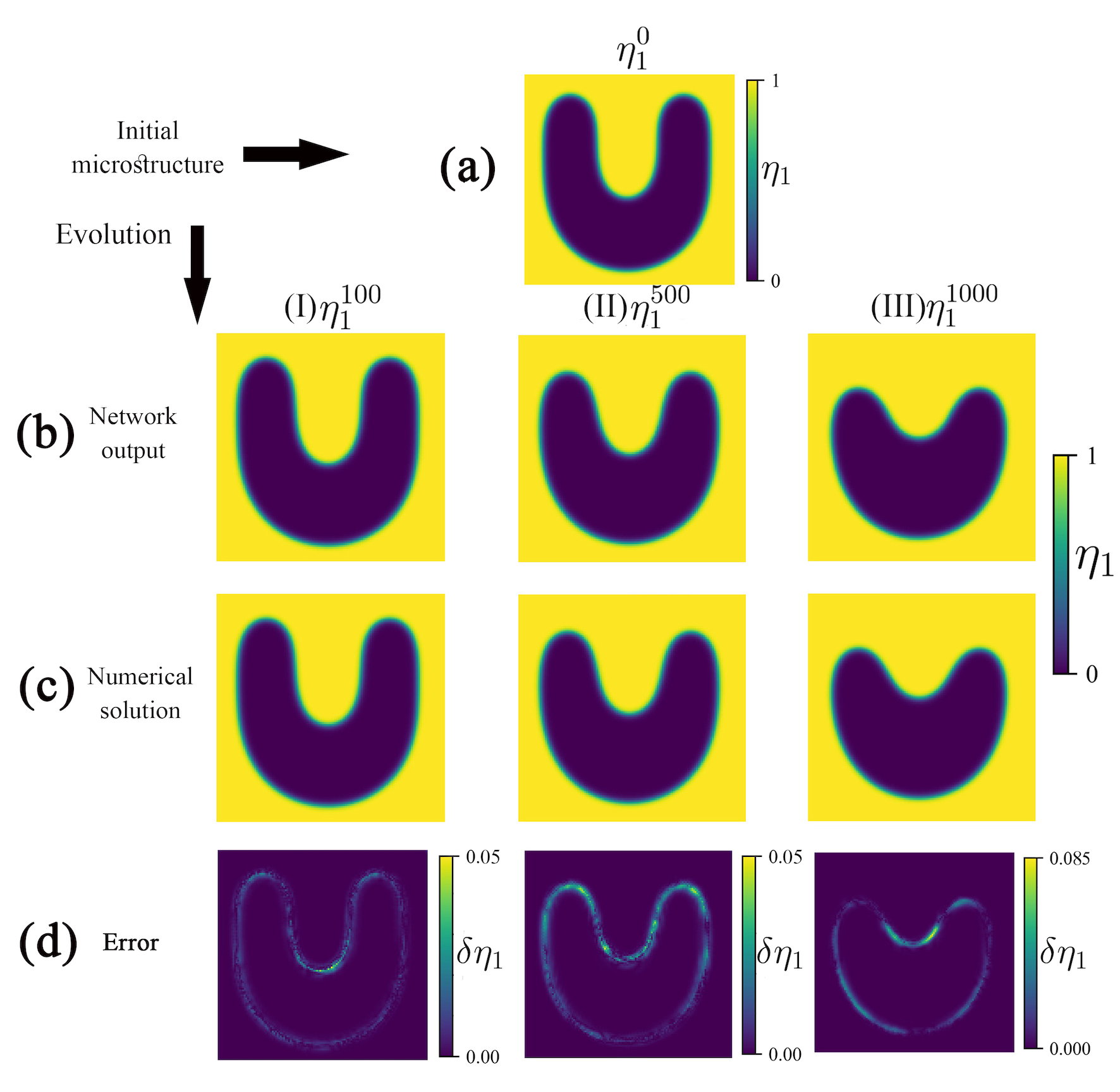}
    \caption{\textbf{Comparison of network prediction for different timesteps forward.} Three neural networks, trained to reproduce 100 (I), 500 (II), and 1000 (III) timesteps forward. 
    (a) is the initial configuration of an example microstructure not included in the training dataset. Row (b) is the output of our network, (c) is the result of the numerical solution of the Fan-Chen model, and (d) is the error between (b) and (c).}
    \label{fig:multipletimesteps}
\end{figure}

\begin{figure}[H]
    \centering
    \includegraphics[width=0.85\textwidth]{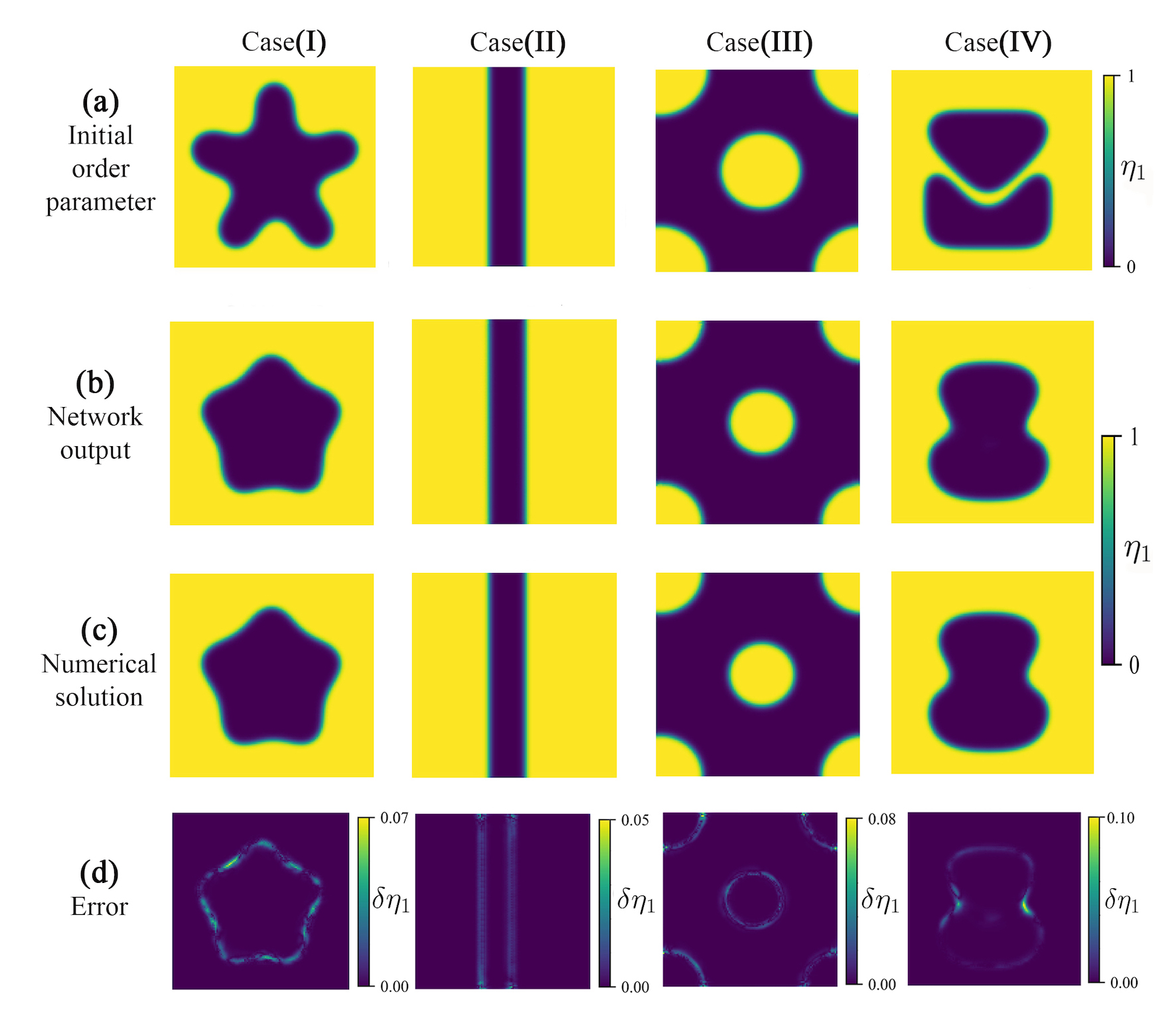}
    \caption{\textbf{The neural network output for different initial structures.} The U-net output compared to the results of the numerical simulation of the Fan-Chen model for different initial order parameter configurations. Row (a) shows four examples of the order parameter ($\eta_1$) in time (n), (b) the evolution of (a) after 500 timesteps, output by the neural network. Row (c) is the evolution of (a) after 500 timesteps obtained from the numerical solution of the Fan-Chen model. Row (d) is the error between (b) and (c).}
    \label{fig:4_nextprediction}
\end{figure}

As outlined above, the neural network can be trained to reproduce the solution 
of the Fan-Chen model at any desired timestep, Fig.~\ref{fig:multipletimesteps}
compares three models of the network that are trained to 
reproduce solutions after three different timesteps. 
In this figure, (a) is the initial configuration of the order parameters. Note that the neural network has not seen this shape in the training process. Column (I) shows the microstructure after 100 timesteps, column (II) after 500 timesteps, and column (III) after 1000 timesteps (the timestep size is 0.008). Row (b) displays results from the neural network, row (c) from the numerical solution of the Fan-Chen model, 
and row (d) illustrates the error between the neural network and the numerical solution of the Fan-Chen model. It appears that as we increase the jump forward to the next timestep from 100 to 1000, the error goes up slightly in some particular areas of the microstructure, namely, at the grain boundaries, where the order parameters change from 1 to 0 and vice versa. However, the mean absolute error for three cases remains close (0.0011 for I, 0.0015 for II and 0.0021 for III). Furthermore, the relative error in total free energy for all three cases is around 0.002. We can conclude from this that choosing larger steps in the order parameter calculation 
has a limited impact on the error or overall performance of the neural network. In the rest of this work, we test the trained network to reproduce 500 timesteps forward with other microstructures, specifically with those that had not been included in the training dataset. Fig.~\ref{fig:4_nextprediction} depicts four examples of order parameter  
output after 500 timesteps forward. 

Fig.~\ref{fig:4_nextprediction} (a) shows the configuration of the $\eta_1$ order parameter at timestep $n$ ($\eta_1^{n}$). The evolution of (a) after 500 timesteps (i.e., $\eta_1^{n+500}$) is shown in (b) and (c) as output by the trained neural network and by the numerical solution of the Fan-Chen model, respectively. Row (d) shows the error between the results of the neural network and the numerical solution of the Fan-Chen model. Evidently, those output by the network agree well with those from the numerical solution of the Fan-Chen model. We also calculate the free energy of both cases, (b) and (c) through equation \ref{eq_free_en}. The results are presented in table \ref{tab:table}. 

\begin{table}[H]
\centering
\caption{Free energy of the system for the cases shown in Fig. \ref{fig:4_nextprediction}. The mean absolute error (MAE) and the mean relative error (MRE) in 
the free energy for each case is also listed.}
    \begin{tabular}{c|c|c|c|c}
    \hline
        Cases                & I        & II     & III &IV\\
    \hline
     Free Energy, Fan-Chen model    &-2750.24  &-3328.13& -1258.49    &-2771.49\\
    Free Energy, trained network& -2747.93 &-3326.04& -1254.87    &-2767.07 \\
     MAE                & 2.311    & 2.085  &  -3.62    &4.42 \\
     MRE (\%)             &0.08    & 0.06  &   0.28    &0.15\\
    \hline      
\end{tabular}
\label{tab:table}
\end{table}

The case (I) in Fig. \ref{fig:4_nextprediction} shows a grain inside another grain. Such topological scenarios are straightforward for the neural network to reproduce since just two grains exist, no box boundaries are involved, and the area occupied by grain number 1 (yellow) is larger than that of grain number 2. The neural network handles large curvatures (translating to a large capillary driving force in the 
Fan-Chen model) in this microstructure very well. Although the error in 
the order parameter in some areas of the microstructure reaches 0.07, the mean absolute error remains below  0.0013 over the whole domain. The mean relative error in the free energy for this case is 0.08\%, showing that the network's performance in such cases is excellent. 

Case (II) in Fig. \ref{fig:4_nextprediction} shows a microstructure in which the area of grain 2 remains constant, and the microstructure does not change over time. This case shows that the neural network recognizes this special case very well. The maximum absolute error in the order parameter is 0.05, and the relative error on the free energy is 0.06\%.

The third example, i.e., case (III) in Fig. \ref{fig:4_nextprediction} is another special case in which multiple grains with the same order parameter are inside each other. In this scenario, the grain with the lowest area shrinks faster than the others. The absolute error in some areas near the box boundary reaches 0.08, while in the grain boundaries, it is less than 0.04 and in the whole domain, the mean absolute error is below  0.0016. The value of the relative error on the free energy is 0.28\%, which indicates that the neural network also outputs the microstructures accurately. 

The fourth example,  case (IV) in  Fig. \ref{fig:4_nextprediction}, is a scenario with two grains close to each other with concave and convex boundaries, respectively. In this configuration, these close grains merge instead of shrinking individually. Even though the absolute error in some points reaches 0.10, in other areas of the grain boundary, it is less than 0.02. The mean relative error of 0.15\% for the free energy indicates that the neural network can accurately reproduce the evolution of the order parameter in this case as well.

In addition to the special cases discussed above, we also made comparisons for all microstructures in the test dataset consisting of 2000 different microstructures. The mean absolute error in the order parameter in our test dataset is less than 0.003. However, at the boundaries of the microstructures, it might reach around 0.1 in a few points. The relative error in the free energy can reach 1.0\% in a few special cases, while it remains below  0.1\% for most cases in the test dataset. This shows that the trained neural network output for the order parameter evolution is quite accurate after 500 timesteps. Next, we investigate the application of the same neural network recursively to reproduce the whole time series of the order parameters evolution.

\subsection{Neural network output for time series}

\begin{figure}[H]
    \centering
    \includegraphics[width=0.9\textwidth]{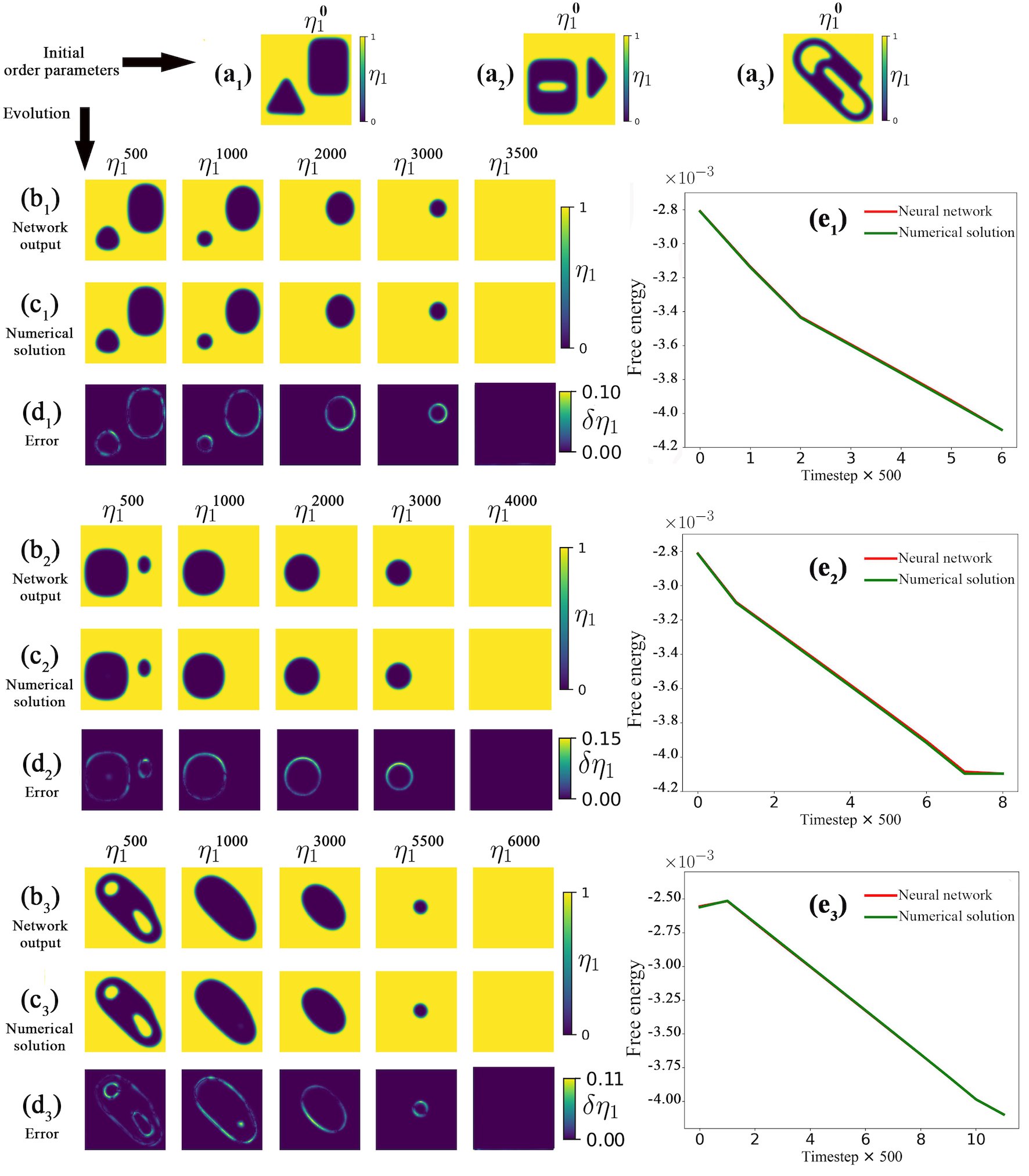}
    \caption{\textbf{The neural network output for time series.} Application of the trained neural network for the evolution of the order parameters (a$_i$) $i=1,2,3$: The initial configuration of the order parameter ($\eta_i^{0}, i=1$), (b$_i$) $i=1,2,3$: the evolution of the initial order parameter calculated by numerical solution of the Fan-Chen model, (c$_i$) $i=1,2,3$: the evolution of the initial order parameter output 
    by the neural network,
    (d$_i$) $i=1,2,3$: the error, 
    (e$_i$) $i=1,2,3$: the changes of free energy in time after every 500 timesteps from the trained network and numerical solution of the Fan-Chen model.}
    \label{fig:multi_prediction}
\end{figure}

The neural network outputs the next desired timestep with reasonable accuracy, as described in the previous section. The network's output can be fed again to the input recursively to reproduce the order parameter's whole time evolution. Fig. \ref{fig:multi_prediction} shows the temporal evolution of the three different microstructures according to the Fan-Chen model. We choose here grain microstructures different from those used in the training and testing process.

Fig. \ref{fig:multi_prediction} (a$_1$) shows the initial condition of a conventional  microstructure, containing two grains. (b$_1$) shows the evolution of (a$_1$) as calculated by the neural network. (c$_1$) shows the microstructure evolution of the starting configurations (a$_1$), calculated by solving the PDEs of the Fan-Chen model numerically (as explained in the Methods section), where $\eta_1$ is shown after 500, 1000, 2000, 3000, and 3500 timesteps. The error in the trained network output is depicted in (d$_1$). The result shows that the error accumulates slowly after each prediction, and at the end of the simulation, it goes up to 0.1 in some of the points in the domain, while the mean absolute error during the simulation drops from 0.0017 to 0.0010. (e$_1$) also compares the total free energy of the domain, calculated based on the gradients of the order parameters at the interfaces. The error accumulation in the total free energy is quite small, reaching only 0.17\% during the simulation. This result indicates that the accuracy of the network output is still quite high for time series prediction. 

The second example, as shown in Fig. \ref{fig:multi_prediction} (a$_2$), considers two grains in the domain as in the preceding example, but this time one of them contains a secondary smaller grain. The temporal evolution of this scenario obtained from the neural network output and the numerical solution of the Fan-Chen model is shown in (b$_2$) and (c$_2$), respectively. (d$_2$) shows the error, which in this case increases up to 0.15 at the end of the simulation. (e$_2$) shows the free energy of the microstructure, calculated based again on the transition regions of the order parameters marking the interface regions. The maximum relative error on the free energy accumulating during the simulation between the two methods is 0.25\%. 

The last case (a$_3$) in Fig. \ref{fig:multi_prediction} shows an initial configuration of a complicated microstructure far from the dataset used to train the neural network. As described in the previous section, first, the grain with the smaller area shrinks and disappears, and then the grain with the largest area absorbs the other one. Both methods reproduce this trend correctly, as shown for the neural network output in in (b$_3$) and the numerical solution in (c$_3$). The errors in the order parameter (d$_3$) and in the free energy (e$_3$) illustrate that the neural network is able to reproduce the evolution of this microstructure scenario (a$_3$), which was not included in the training dataset. 

\subsection{Computational speed}

The results shown above demonstrate a good neural network accuracy for the time evolution of the order parameters for a wide variety of microstructural topologies. To investigate the increase in computational speed obtained with the network relative to the conventional forward simulation, 
we employ one core of an Intel(R) Core(TM)i5 CPU, clocked at 3.2 GHz, for 
both numerical solution of the Fan-Chen model and neural network output. 
The neural network calculates the next frame (500 timesteps forward) in 38 milliseconds (measured on average over multiple evaluations), while the FFT solver computes 500 timesteps in 3.448 seconds, meaning that in this case, the neural network calculates the next frame around 90 times faster than the FFT based solver. Note that the computation time for evaluating the microstructure evolution by the trained network is independent of the initial configuration of the order parameters. In contrast, the convergence of the numerical solver highly depends on the complexity of the order parameter distribution. Furthermore, the speedup reported above is merely an estimate. Both, the neural network and the FFT-based calculations could be performed in more efficient ways. For example, here, we focused mainly on output 500 timesteps 
forward. However, as shown in Fig. \ref{fig:multipletimesteps}, with the same amount of calculations, the neural network will also work for output 1000 timestep forward. Similarly, the FFT-based solver could employ advanced adaptive time integration algorithms to speed up the calculations. 

\section{Conclusions}
\label{conc}

For a broad range of physical, biological, and chemical systems, 
the phase-field model provides a powerful and versatile computational 
methodology suited for calculating the evolution of microstructures 
and their associated properties. Current phase-field models and numerical methods, however, are computationally expensive by nature, necessitating high-performance computing systems as well as complex numerical integration approaches to obtain a reasonable level of precision. In this study, we presented a fast, efficient and accurate way to reproduce solutions of the 
Fan-Chen model for grain growth. 
For this goal, a dataset of non-conserved structural order parameters that define the grain microstructures and their evolution was generated by numerical simulation. A U-net architecture was then used to learn from a subset of these conventional full-field solutions the time-dependent evolution of the patterns (i.e. microstructures) in the dataset. 
The U-net is a Convolutional Artificial Neural Network architecture that can receive the matrix of order parameters and extract important features in the latent space while transferring the spacial information through skip connections.
The trained U-net is then evaluated with the use of the test dataset as well as with special microstructure topology cases that had not been included in the original training dataset. Our findings show that the U-net is able to find the solution to the time-dependent PDE (here Fan-Chen model) by mapping the order parameter in the next desired timestep to the previous one. We trained the U-net on the simulation data with frames generated each 100, 500, and 1000 timesteps. Although larger leaps might be possible, we selected 500 timesteps as a reasonable compromise between error and efficiency. Recursive application of the neural network results in generating the whole time series of the order parameter evolution with minimum accumulated error in our case. Besides the reasonable accuracy of the method, the U-net is shown to be about 90 times faster than the numerical simulation with a mean absolute error of the order parameter and a mean relative error of free energy around 0.0015 and 0.002, respectively. Besides, the solution from the U-net surrogate model can be fed directly into a conventional high-fidelity phase-field model and evolve by numerical simulation to jump in time and speed up the conventional high-fidelity phase-field simulation, opening the opportunity for highly efficient hybrid modeling architectures. We can take advantage of this accelerated framework to quickly explore a large phase-field input domain for all kinds of mesoscale problems that can be described by non-conserved order parameters and the Fan-Chen equation. 
Future studies on multi-scale and multi-physics simulations will benefit from these findings.

\vspace{8mm}
\noindent
\textbf{Data availability}: The data that support the findings of this study are available from the corresponding author upon reasonable request.
\vspace{4mm}

\noindent
\textbf{Code availability}: 
The code used for machine learning in this study is open-source and accessible in \href{https://github.com/tensorflow}{github.com/tensorflow}. Other codes and scripts used in this work are available from the corresponding author upon reasonable request.
\vspace{4mm}

\noindent
\textbf{Acknowledgement}:
The support from Shahid Beheshti University is gratefully acknowledged. This research did not receive any specific grant from funding agencies in the public, commercial, or not-for-profit sectors.
\vspace{4mm}

\noindent\textbf{Author contributions}: I.P., J.R.M. and G.A. developed the initial concept and workflow. I.P. and J.R.M developed the PF and the neural network framework. N.H.S. and R.G. contributed to the neural network design and training approach. I.P. and carried out the PF calculations and neural network training. D.R. , B.S. and G.A. helped in the interpretation of the results. I.P. and J.R.M prepared the initial draft of the manuscript. J.R.M supervised the work. All authors discussed and contributed to preparing the final version of the manuscript.
\vspace{4mm}

\noindent\textbf{Competing interests}:
The authors declare no competing financial or non-financial interests.

\bibliographystyle{naturemag}
\bibliography{Short,Refs}
\end{document}